# Absolute and Relative Bias in Eight Common Observational Study Designs: Evidence from a Meta-analysis

Jelena Zurovac[a]*, Thomas D. Cook[b], John Deke[a], Mariel M. Finucane[a], Duncan Chaplin[a], Jared S. Coopersmith[a], Michael Barna[a], and Lauren Vollmer Forrow[c]

[a]Mathematica, Inc, Washington, DC

[b]George Washington University, Washington, DC

[c] Independent consultant, Oxford, United Kingdom

*Jelena Zurovac is a Senior Researcher, Mathematica, Inc., Washington, DC 22202 (email: jzurovac@mathematica-mpr.com); Thomas D. Cook is a Research Professor, George Washington university, Washington, DC 20052 (t-cook@northwestern.edu), John Deke is a Senior Fellow, Mathematica, Inc, Princeton, NJ 08543; Mariel M. Finucane is a Principal Researcher, Mathematica, Inc., Cambridge, MA 02139; Duncan Chaplin is a Principal Researcher, Mathematica, Inc., Washington, DC 20052; Jared S. Coopersmith is a Senior Researcher, Mathematica, Inc., Washington, DC 20052; Michael Barna is a Researcher, Mathematica, Inc., Ann Arbor, MI 48104; Lauren Vollmer Forrow is Independent Consultant, Oxford, United Kingdom.

# Absolute and Relative Bias in Eight Common Observational Study Designs: Evidence from a Meta-analysis

Observational studies are needed when experiments are not possible. Within study comparisons (WSC) compare observational and experimental estimates that test the same hypothesis using the same treatment group, outcome, and estimand. Meta-analyzing 39 of them, we compare mean bias and its variance for the eight observational designs that result from combining whether there is a pretest measure of the outcome or not, whether the comparison group is local to the treatment group or not, and whether there is a relatively rich set of other covariates or not. Of these eight designs, one combines all three design elements, another has none, and the remainder include any one or two. We found that both the mean and variance of bias decline as design elements are added, with the lowest mean and smallest variance in a design with all three elements. The probability of bias falling within 0.10 standard deviations of the experimental estimate varied from 59 to 83 percent in Bayesian analyses and from 86 to 100 percent in non-Bayesian ones -- the ranges depending on the level of data aggregation. But confounding remains possible due to each of the eight observational study design cells including a different set of WSC studies.

# 1. INTRODUCTION

Evidence-based policy making seeks to identify effective social programs whose dissemination is likely to improve social welfare (Lawson 2016; Baron 2018; Gamoran 2019). Randomized control trials (RCTs) are best for this because they are unbiased in expectation and the standard error of ther impact estimate reflects the probability of mis-estimating effects. But RCTs are not always possible or appropriate (Heckman and Smith 1995; Heckman 2020), and so observational studies are needed that compare outcomes between non-randomly formed treatment and comparison groups, a non-random selection process that is usually related to confounding factors correlated with the outcome. If these confounds were fully known, perfectly measured, and appropriately used as covariates, then the bias they otherwise cause is ruled out in theory. But meeting these conditions with individual observational studies is well-nigh impossible, though exceptionally large and heterogeneous datasets -- such as Scandinavian registries or Facebook data (Eckles and Bakshy 2021) -- can provide a reasonable multi-dimensional approximation. In less data-rich contexts, however, causal inference is less secure and has historically led to many controversies about specific causal claims. We assume fewer controversies would arise if researchers knew, and could promote, observational study designs that reduce the risk of bias to tolerable levels.

      This paper highlights a way to test this possibility when access to covariate data is far from unlimited. It relies on the "within study comparison" (WSC) method that compares observational and RCT estimates that test the same hypothesis using the same treatment group, outcome and estimand (LaLonde 1986; Cook et al. 2008). Bias is then defined as the difference between the RCT and observational study estimates. Using this method, we test the bias in quasi-experimental designs composed of up to three

design elements that past meta-analyses of WSC applications have shown to reduce bias when used alone (Glazerman et al. 2003; Bloom et al. 2005; Smith and Todd 2005). One is a local comparison group from the same market for goods and services as the treated group. The expectation is that this will make the treatment and comparison groups as similar as possible at baseline, including on unobservables like local policies and practices that might influence outcomes. The second design element is the pretest, a pre-intervention measure of the outcome that is usually more highly correlated with the outcome than other baseline measures and that is also often correlated with selection into treatment (Campbell and Stanley 1963; Hedges and Hedberg 2007). The third element is a "relatively rich" set of other covariates from multiple domains in hopes they will account for additional confounders unrelated to the pretest and local comparison group. We use the label "relatively rich" because many fewer covariates are involved than with "big data" that are arguably truly rich. This tri-partite approach to bias control is especially relevant since many quasi-experimental applications can manipulate some, or even all, of these three elements.

    Our study estimates bias in eight observational study designs. These result from factorially crossing two values of each of the three design elements. Of the eight, the most sophisticated combines a local comparison group, a pretest and other rich covariates; the least sophisticated uses a non-local comparison group, no pretest, and no other covariates; and the six other alternatives combine any one or any two of the elements. Although previous WSC reviews have shown that each element lowers bias when examined singly, none has estimated the mean and variance of bias over all eight design cells. Variance is key here, for it helps little to know that the average bias is close to zero over a set of WSCs if the bias in most single WSCs is far from zero. Other meta-analyses of WSC studies in this series have estimated mean bias in regression-

discontinuity (Chaplin et al. 2018) and in comparative interrupted time-series (Coopersmith et al. in press) studies. While each presented the distribution of study-specific bias means to describe how tight it was over all the WSCs examined, neither directly examined the variance of bias estimates within a single WSC.

Estimates of mean bias and its variance are subject to random error. This requires us to specify a region of practical equivalence (ROPE) that describes the level of bias we will consider as functionally equivalent to being unbiased (Wilde and Hollister 2007). We adopt a ROPE of bias within 0.10 standard deviations of the RCT estimate, equivalent in education to raising academic achievement from the median to the 54th percentile. But we also report bias means and standard deviations so that readers can apply their own ROPE standard of tolerable bias.

## 2. DATA

### *2.1. Identifying studies.*

To identify WSCs for inclusion in this meta-analysis we (1) searched electronic journal databases with keywords: "within study comparison", "design replication study", "RCT benchmark", and "internal replication study," (2) reviewed references in identified WSCs and WSC reviews (e.g., Villar and Waddington 2019), and (3) contacted authors of known WSCs. This process ended in September 2019 and identified 95 published and unpublished manuscripts. We screened out 26 because of no RCT, no comparison group in the observational study, or an intervention that was not about social policy. We screened out 20 more because of confounding factors in the WSC, including small overlap between the RCT and observational study treatment groups, lack of RCT balance on the pretest or most baseline variables, or different estimands in the RCT and observational study. We excluded five other WSCs because information was missing for estimating bias or determining how well the RCT was implemented. Of 50 remaining WSCs, 11 did not report standard errors of the RCT and/or observational study. Thus, 39 WSCs remained in the meta-analysis.

*2.2. Study review and coding*

Studies were reviewed in several stages. First, a research assistant recorded information guided by a comprehensive protocol. Then, an author of this manuscript provided feedback. After a re-review by another manuscript author, the three parties met to reconcile remaining discrepancies. Many studies reported multiple bias estimates for different outcomes, subgroups, sites, waves of posttest measurement, and ways of analyzing the data. We call each such bias estimate "a contrast". We excluded contrasts for subgroups if an overall finding was reported, as well as contrasts that were reported only in appendices or that employed only marginally different methods than the main analyses. Where a summary index was presented plus its constituent measures—say, total income and then separate wage and non-wage income—we used only the summary. With multiple posttest data waves, we used just the first and last.

*2.3. Sample*

Of the 39 WSCs, 27 were published in peer-reviewed journals, one in a non-peer-reviewed journal, and 11 were working papers or reports. The interventions were heterogeneous, though 24 addressed academic achievement in education. The rest were in job training (9), welfare (3), labor (3), health (2), environmental policy (2), and migration (1). Nineteen WSCs varied use of a pretest, 8 use of a local comparison group, and 11 use of rich covariates. In every WSC, baseline RCT treatment and control differences was consistent with chance.

*2.4. Data Structure and its Consequences for Analytic Strategy*

Our data have a crossed hierarchical data structure: 26 of the 39 studies estimated bias for multiple design cells and 21 of the 39 studies included multiple outcome domains. To account for this structure, we use both Bayesian and non-Bayesian hierarchical meta-analytic regression models. The 39 WSCs use only 25 RCT datasets as benchmarks, but this level of clustering could not be accounted for since models with this additional hierarchy could not converge. The generic trade-off between precision and meta-bias (bias in the bias estimates) could not be adequately addressed in a single analysis, and we later present details of a mixed strategy that involved using a Bayesian analysis to increase precision by borrowing information over studies and over the eight observational study design cells. Since borrowing over cells can make the bias estimates in one cell dependent on those in another, the possibility of meta-bias arises. To counter this, we also conducted a non-Bayesian analysis that analyzed the eight design cells independently and so no borrowing was involved. Of course, the two models differ in other ways too, including the target population. The non-Bayesian work accounts for random variation in the mean bias across studies, while the Bayesian model further accounts for random variation in the within-study distribution of bias. As a result, the Bayesian work postulates more uncertainty and provides more cautious estimates of the probability of future bias from the 39 WSC studies examined here. However, they were sampled without reference to a well-specified formal population, and education studies of academic achievement predominate in the sample.

The WSCs vary in the observational study designs to which they contribute bias estimates. Three studies contribute to four cells, four to three cells, 19 to two, and 13 to a single cell. Therefore, the composition of studies within each design cells differs in terms of the populations, outcomes and settings studied, as well as in the treatment and outcome variants and in how the observational analysis we conducted. Covariates were used to address these selection differences by design cell, but we see below that they do not directly speak to study differences in human populations, settings, times, interventions, outcomes, or methods of observational study analysis.

## 3. METHODS

### 3.1. Definition of key variables

An estimate of observational study bias y is defined as the difference between the observational study and RCT impact estimates in standard deviation units.

$$y_{ijkl} = \frac{\hat{\theta}_{ijkl}^{(Obs)} - \hat{\theta}_{ijkl}^{(RCT)}}{S_{ijkl}^{(RCT)}} \quad (1)$$

for contrast $i$, design $j$, study $k$, and outcome domain $l$. $\hat{\theta}^{(Obs)}$ and $\hat{\theta}^{(RCT)}$ are the impact estimates from the observational study and the RCT respectively, and $S^{(RCT)}$ is the standard deviation of the outcome for the RCT treatment and control groups. If a pooled standard deviation was not reported, we approximated it (see Supplement A). Bias estimates are positive when the observational study impact estimate is higher in the direction of the socially desired outcome the RCT seeks to achieve, indicating that the observational study over-estimates intervention effectiveness.

We approximate $s_{ijkl}^{(y)}$, the standard error of standardized bias estimate $y_{ijkl}$, as:

$$s_{ijkl}^{(y)} = \frac{1}{S_{ijkl}^{(RCT)}} \sqrt{\left(\left(s_{ijkl}^{(Obs)}\right)^2 + \left(s_{ijkl}^{(RCT)}\right)^2\right) \left( \frac{\frac{1}{N_{ijkl}^{(Contr)}} + \frac{1}{N_{ijkl}^{(Comp)}}}{\frac{1}{N_{ijkl}^{(Contr)}} + \frac{1}{N_{ijkl}^{(Comp)}} + \frac{2}{N_{ijkl}^{(Trt)}}} \right)} \quad (2)$$

where $S^{(RCT)}$ is the pooled outcome standard deviation from the RCT, and $s^{(Obs)}$ and $s^{(RCT)}$ are the standard errors of $\hat{\theta}^{(Obs)}$ and $\hat{\theta}^{(RCT)}$, respectively. $N^{(Contr)}$, $N^{(Comp)}$, and $N^{(Trt)}$ are sample sizes for the RCT control, observational study comparison, and shared treatment groups. The $2/N^{(Trt)}$ term accounts for covariance between the impact estimates from the observational study and RCT, which is due to these impact estimates sharing a common treatment group. See Supplement A for the derivation of equation (2).

In this study, an observational impact estimate is considered to account for a pretest ($P_j = 1$) if one or more pre-intervention outcomes are included in matching analyses or as covariates in a regression. Also included was use of a proxy pretest, a variable that measures the same underlying concept as the outcome – for example, a good proxy pretest for SAT math scores might be a state standardized math assessment. A comparison group is considered local ($L_j = 1$) if all comparison units are in the same market for goods and services as the treated units. For educational interventions, this meant that all the comparison and treatment units were in the same school district. Covariates used in an observational study were considered rich ($R_j = 1$) if a study matched on, or regression-adjusted for, variables in at least four unique conceptual domains, which we defined prior to analysis (See Supplement A). The most common covariate domains were basic demographics (age, gender, race), educational characteristics, socioeconomic characteristics, and geographic characteristics.

We used four study characteristics, denoted $X_k$, to try to account for differences in the composition of studies within each design cell: (a) whether the RCT intervention is in education, (b) whether a given study includes more design cells than most other studies, (c) the number of contrasts in each study, and (d) whether study authors were, or had been, part of the research team conducting WSCs at Northwestern University. These are not likely to account for all ways in which study populations, outcomes, settings, etc. vary within and between the eight observational study designs.

*3.2. Analyzing data at multiple levels*

We examine bias at three levels. Contrast-level bias is based on one bias estimate for each subgroup in the analysis, whether it reflects different human populations, settings, times or representations of the cause and of the effect. The analyses we present include 473 such contrasts, from 1 to 46 per study. While this level offers the largest sample size of the three we consider, it is the most sensitive to sampling error from smaller subgroups and less reliable sub-test measures. *Hypothesis-level* bias is estimated as a within-study average of all contrast-level estimates within an outcome domain – e.g., "How did the intervention affect employment for the full study sample?" as opposed to "those in different locations or for different respondent groups?" (Friedlander and Robins 1995)?" Over the 39 studies in our sample there are 65 hypothesis-level bias estimates, illustrating how few studies sought to affect multiple endpoints, about 1.7 per study. Study-level bias is the most reliable and has been most used in past syntheses of WSC results (Chaplin et al. 2018; Coopersmith et al. in press). It requires averaging over all the contrast-level estimates in a WSC. But in the present context it has two potential disadvantages. First is the low number of studies within most pf the eight design cells -- from 3 to 17 with a median of 11; and second, average study-level bias near zero may obscure hypothesis-level bias estimates with different signs that are the product of valid under- and over-estimates of bias rather than random error. The present sample has only 1.7 hypotheses per study, though; and several WSCs have explored the possibility of heterogeneity in such bias without finding evidence for it (Cook et al. 2020; Brown et al. 2021). Nonetheless, the potential for mis-estimated aggregation indicates we should rely more on the *hypothesis level* that best threads the needle between the unreliability of contrast-level estimates and the possible treatment heterogeneity of study-level estimates.

### 3.3. Bayesian hierarchical meta-analytic distributional regression model

We model $y_{ijkl}$ — the standardized bias estimate from contrast $i$, design $j$, study $k$, and outcome domain $l$ – as follows.

$$y_{ijkl} = \mu_j + a_k + b_l + c_{jk} + d_{jl} + e_{kl} + f_{jkl} + g_i + \beta X_k + \varepsilon_{ijkl} \qquad (3)$$

Equation (3) defines the standardized bias estimate $y_{ijkl}$ to be comprised of ten effects that allow us to estimate the distinct sources of variability that contribute to each bias estimate. $\mu_j$ is the mean bias of design $j$ – the components of $\mu_j$ are described in Section 3.3.1. $a_k$, $b_l$, …, $g_i$ are random effects of study $k$, outcome domain $l$, …, and contrast $i$, respectively – these random effects and their variance regressions are described in Section 3.3.2. The $\beta$s are effects of characteristics $X_k$ of study $k$, defined in Section 3.1. We hold these study characteristics constant while comparing across designs, using the marginal effects approach described in Section 3.3.3. Lastly, $\varepsilon_{ijkl}$ is a normally distributed error term – its variance/covariance structure is described in Section 3.3.4. We fit the model using the probabilistic programming language Stan (Carpenter et al. 2017), using its R interface rstan.

### 3.3.1. Mean bias (μ)

We model $\mu_j$, the mean bias of design $j$, as follows.

$$\mu_j = \theta + \gamma_P P_j + \gamma_L L_j + \gamma_R R_j + \delta_j \qquad (4)$$

In Equation (4), $\theta$ is the overall mean bias across designs. $\gamma_P$, $\gamma_L$, and $\gamma_R$ are the effects of the three design elements of interest: pretest ($P_j$), local comparison group ($L_j$), and rich covariates ($R_j$). Lastly, $\delta_j$ captures interactions between the design elements. For example, adjusting for a rich set of covariates may be less crucial for unbiasedness in a study that uses a local comparison group than in a study that does not use a local comparison group.

*3.3.2. Random effects (a, b, ..., g)*

Our model includes random effects at the study and outcome domain levels ($a$ and $b$, respectively), for all two- and three-way combinations of design, study, and outcome domain ($c$, $d$, $e$, and $f$ – for example, $c$ is a random effect of design-by-study interactions), and at the contrast level ($g$). To allow for the existence of bias outliers (above and beyond what can be explained by the sampling variability reflected in the bias estimates' standard errors), we model the random effects as coming from Student $t$ distributions, with degrees of freedom at each level of the hierarchy $\nu_a, \nu_b, \ldots, \nu_g$ estimated using a $Gamma(2, 0.1)$ prior. We impose a lower bound of 2.5 on each $\nu$ parameter ($\nu_a, \nu_b, \ldots, \nu_g$) to ensure a finite variance of each batch of random effects ($a$, $b$, ..., $g$).

We use variance regression methods to flexibly estimate variation in bias (represented in the form of standard deviations in equation 5), adjusting for study characteristics $X_k$. We use the familiar log-linear form to allow the influence of design elements and other covariates to be expressed as a linear combination, while restricting the standard deviations to be positive.

$$
\begin{aligned}
a_k &\sim t_{\nu_a}\left(0, \sigma_k^{(a)}\right) & \ln \sigma_k^{(a)} &= \alpha^{(a)} + \zeta^{(a)} X_k \\
b_l &\sim t_{\nu_b}\left(0, \sigma^{(b)}\right) & \ln \sigma^{(b)} &= \alpha^{(b)} \\
c_{jk} &\sim t_{\nu_c}\left(0, \sigma_{jk}^{(c)}\right) & \ln \sigma_{jk}^{(c)} &= \alpha^{(c)} + \zeta^{(c)} X_k + \varphi_P^{(c)} P_j + \varphi_L^{(c)} L_j + \varphi_R^{(c)} R_j + \xi_j^{(c)} \\
d_{jl} &\sim t_{\nu_d}\left(0, \sigma_j^{(d)}\right) & \ln \sigma_j^{(d)} &= \alpha^{(d)} + \varphi_P^{(d)} P_j + \varphi_L^{(d)} L_j + \varphi_R^{(d)} R_j + \xi_j^{(d)} \quad (5)\\
e_{kl} &\sim t_{\nu_e}\left(0, \sigma_k^{(e)}\right) & \ln \sigma_k^{(e)} &= \alpha^{(e)} + \zeta^{(e)} X_k \\
f_{jkl} &\sim t_{\nu_f}(0, \sigma_{jk}^{(f)}) & \ln \sigma_{jk}^{(f)} &= \alpha^{(f)} + \zeta^{(f)} X_k + \varphi_P^{(f)} P_j + \varphi_L^{(f)} L_j + \varphi_R^{(f)} R_j + \xi_j^{(f)} \\
g_i &\sim t_{\nu_g}\left(0, \sigma_{j[i]k[i]}^{(g)}\right) & \ln \sigma_{jk}^{(g)} &= \alpha^{(g)} + \zeta^{(g)} X_k + \varphi_P^{(g)} P_j + \varphi_L^{(g)} L_j + \varphi_R^{(g)} R_j + \xi_j^{(g)}
\end{aligned}
$$

In Equation (5), $\alpha^{(a)}, \alpha^{(b)}, \ldots, \alpha^{(g)}$ are the intercepts of the variance regressions at the study, outcome domain, …, and contrast levels, respectively, describing the baseline amount of bias variation at each level. The $\zeta$s describe the effects of study characteristics $X_k$ on the random effects' variances at each level. The $\varphi$s (and $\xi$s) are effects of design elements $P_j$, $R_j$, and $L_j$ (and their interactions) on the random effects' variances at each level. The hierarchical priors for the $\alpha$s, $\zeta$s, $\varphi$s, and $\xi$s are described in section 3.3.5.

*3.3.3. Adjusting for study characteristics*

To increase the validity of comparisons across designs, we estimate what the variance of bias would have been for each design if it had been applied to all contrasts in the sample. That is, when comparing the variance of bias across designs, we hold the study characteristics $X_k$ constant. To do this we begin by calculating eight values of the mean and variance of bias for every contrast in our sample using each contrast's observed values of $X_k$. We calculate the first of the eight values of mean and variance of bias by setting the design elements (and therefore their interactions) equal to those of the first design, we calculate the second values by setting the design elements equal to those of the second design, and so on. For each design, we then marginalize over our sample's variance of study characteristics. Specifically – for each design – we take an average over contrasts of the mean bias values and the variance of bias values.

*3.3.4. Correlated error terms (ε)*

The error terms $\varepsilon$ in equation (3) have a multivariate normal distribution with a block-diagonal variance-covariance matrix. Each block represents a study; diagonal entries are the squared standard error of the bias estimate, $(s^{(y)})^2$, while off-diagonal entries represent covariance among the estimation errors of contrasts in the same study, with correlation $\rho = 0.8$. Computational tractability led us to use this fixed value rather than estimating $\rho$. However, we tested the sensitivity of our conclusions to this choice and found them to be robust. We assume that the error correlation among contrasts from different studies is zero.

*3.3.5. Hierarchical priors*

We use hierarchical priors to make estimation possible where data are sparse and to improve estimates that are based on richer data.

$$\begin{aligned}\beta &\sim N(0, \tau_\beta) \\ \gamma &\sim N(\phi_\gamma, \tau_\gamma) \\ \delta &\sim N(0, \tau_\delta) \\ \alpha^{(a)}, \alpha^{(b)}, \ldots, \alpha^{(g)} &\sim N(\phi_\alpha, \tau_\alpha) \\ \zeta^{(a)}, \zeta^{(c)}, \ldots, \zeta^{(g)} &\sim N(0, \tau_\zeta) \\ \varphi^{(c)}, \varphi^{(d)}, \ldots, \varphi^{(g)} &\sim N(\phi_\varphi, \tau_\varphi) \\ \xi^{(c)}, \xi^{(d)}, \ldots, \xi^{(g)} &\sim N(0, \tau_\xi)\end{aligned} \qquad (6)$$

Importantly, these priors induce shrinkage across designs in the estimates of the mean and variation in bias. Specifically, the priors allow for arbitrary interactions among the three design elements but "partially pool" towards a simpler additive model in which the three elements all have the same effect. Taking mean bias as an example, recall from Section 3.3.1 that $\mu_j = \theta + \gamma_P P_j + \gamma_L L_j + \gamma_R R_j + \delta_j$. Consider four examples taken from a spectrum of possible degrees of shrinkage across designs. First, if $\tau_\delta$ is estimated to be large, then the model will not shrink across designs – it will estimate mean bias for each design based primarily on the direct evidence from that design, analogous to the "no pooling" non-Bayesian model described in Section 3.4. Second, as the estimate of $\tau_\delta$ approaches zero, the regression equation collapses down towards an additive model: $\mu_j = \theta + \gamma_P P_j + \gamma_L L_j + \gamma_R R_j$. In this case, design cells that share design elements are shrunk towards each other, e.g. designs that use a pretest all share $\gamma_P$. Third, if $\tau_\delta$ and $\tau_\gamma$ are both estimated to be near zero, then the regression equation collapses down further, toward: $\mu_j = \theta + \gamma_P P_j + \gamma_L L_j + \gamma_R R_j = \theta + \phi_\gamma N_j$, where $N_j$ is the number of elements included in the design. In this case, design cells that share the same number of design elements are shrunk towards each other. Fourth and most parsimoniously, if $\tau_\delta$, $\tau_\gamma$, and $\phi_\gamma$ are all estimated to be near zero, then the model moves towards a "complete pooling" model in which we estimate no difference in mean bias across designs: $\mu_j = \theta$. So, through estimation of variance components the Bayesian model allows the data to inform the appropriate degree of pooling. If design differences seem negligible, then the model will pool more and, in our example, more pooling corresponds to smaller estimates of $\tau_\delta$ and $\tau_\gamma$. However, if meaningful heterogeneity in bias is observed across designs, the model will pool less and estimates of $\tau_\delta$ and $\tau_\gamma$ will be larger. Therefore, our estimate of the risk of bias under each design is improved through drawing on information from the other designs, but only to the extent that the data deem them mutually informative and that pooling decisions are not affected by the small cell samples of between 3 and 17 studies. For a more definitive unconfounding of observational study cells and the different sets of studies within them we turn to non-Bayesian analyses.

We use standard $N(0,1)$ priors for all hyperparameters. The only exception is $\phi_\alpha$, which has an improper flat prior. A $N(0,1)$ prior would be anti-conservative in this particular case, in the sense that it would rule out a large negative value of $\phi_\alpha$, which would equate to very small average variance $\exp(\phi_\alpha)$.

### *3.4. The non-Bayesian model*

The non-Bayesian approach following Pustejovsky and Tipton (2021) combines a random-effects meta-regression (implemented using the R package metafor) with cluster-robust variance estimation (CRVE) (implemented using the R package clubSandwich). The multi-level model includes design-specific random effects for each study, hypothesis and contrast, so that each design has its own level-specific estimates of average bias and variation in bias. To obtain standard errors of mean bias estimates that are robust to mis-specification of the multi-level model, we use robust variance estimation, clustering by study.

The specification of the non-Bayesian meta-regression model is:

$$y_{ijkl} = \mu_j + c_{jk} + f_{jkl} + g_i + \beta X_k + \varepsilon_{ijkl} \tag{7}$$

where $c_{jk} \sim N\left(0, \sigma_j^{(c)}\right), f_{jkl} \sim N\left(0, \sigma_j^{(f)}\right)$, and $g_i \sim N\left(0, \sigma_{j[i]}^{(g)}\right)$. The estimation errors $\varepsilon$ are distributed multivariate normal as described in Section 3.3.4. At each level of the model, random effects for different designs are assumed to be mutually independent. We analyzed the model with and without the four covariates, but they made little difference.

We estimated the standard deviation of bias at the hypothesis and contrast levels as:

$$S_j^{\text{hypoth}} = \sqrt{\left(SE(\hat{\mu}_j)\right)^2 + \left(\sigma_j^{(c)}\right)^2 + \left(\sigma_j^{(f)}\right)^2}$$

$$S_j^{\text{contrast}} = \sqrt{\left(SE(\hat{\mu}_j)\right)^2 + \left(\sigma_j^{(c)}\right)^2 + \left(\sigma_j^{(f)}\right)^2 + \left(\sigma_j^{(g)}\right)^2}$$

The variance components of the multi-level working model provide a description of heterogeneity in the true degree of bias at each level of the hierarchy. These prediction distributions get progressively larger at lower levels of analysis, with aggregate bias most stable at the study level and least stable at the contrast level. The predictive distributions are estimated in two ways. First, we calculated a $\beta$-level prediction interval for the bias distribution at the study level by taking

$$\hat{\mu}_s \pm t\left(\nu_s, \frac{1+\beta}{2}\right)\sqrt{\text{Var}(\hat{\mu}_s) + \hat{\tau}_s^2}$$

where $\hat{\mu}_s$ is the estimated average degree of bias in the design subgroup $s$, $\left(\nu_s, \frac{1+\beta}{2}\right)$ is the $\frac{1+\beta}{2}$ quantile of the Student-$t$ distribution with $\nu_s$ degrees of freedom,

Var $(\hat{\mu}_s)$ is the cluster-robust variance estimate of the sampling variance of $\hat{\mu}_s$, $v_s$ is the Satterthwaite degrees of freedom corresponding to Var $(\hat{\mu}_s)$, and $\hat{\tau}_s^2$ is the estimated study-level variance component for subgroup $s$. Prediction intervals for the bias distribution at the hypothesis and contrast levels we calculated similarly as:

$$\hat{\mu}_s \pm t\left(v_s, \frac{1+\beta}{2}\right)\sqrt{\text{Var}(\hat{\mu}_s) + \hat{\tau}_s^2 + \hat{\omega}_s^2}$$

and

$$\hat{\mu}_s \pm t\left(v_s, \frac{1+\beta}{2}\right)\sqrt{\text{Var}(\hat{\mu}_s) + \hat{\tau}_s^2 + \hat{\omega}_s^2 + \hat{\phi}_s^2}$$

where $\hat{\omega}_s^2$ and $\hat{\phi}_s^2$ are, respectively, the estimated hypothesis-level and contrast-level variance components for the design $s$.

We also calculated the proportion of study-, hypothesis- and contrast-level predictive distributions that fall within the range $[-0.1, 0.1]$. We used a normal approximation for these predictive distributions, estimating the proportions as follows:

$$\hat{\pi}_s^{study} = \Phi\left(\frac{0.1 - \hat{\mu}_s}{\sqrt{\text{Var}(\hat{\mu}_s) + \hat{\tau}_s^2}}\right) - \Phi\left(\frac{-0.1 - \hat{\mu}_s}{\sqrt{\text{Var}(\hat{\mu}_s) + \hat{\tau}_s^2}}\right)$$

(Study-level proportion)

$$\hat{\pi}_s^{\text{hypothesis}} = \Phi\left(\frac{0.1 - \hat{\mu}_s}{\sqrt{\text{Var}(\hat{\mu}_s) + \hat{\tau}_s^2 + \hat{\omega}_s^2}}\right) - \Phi\left(\frac{-0.1 - \hat{\mu}_s}{\sqrt{\text{Var}(\hat{\mu}_s) + \hat{\tau}_s^2 + \hat{\omega}_s^2}}\right)$$

(Hypothesis-level proportion)

$$\hat{\pi}_s^{\text{contrast}} = \Phi\left(\frac{0.1 - \hat{\mu}_s}{\sqrt{\text{Var}(\hat{\mu}_s) + \hat{\tau}_s^2 + \hat{\omega}_s^2 + \hat{\phi}_s^2}}\right) - \Phi\left(\frac{-0.1 - \hat{\mu}_s}{\sqrt{\text{Var}(\hat{\mu}_s) + \hat{\tau}_s^2 + \hat{\omega}_s^2 + \hat{\phi}_s^2}}\right)$$

(Contrast-level proportion),

where $\Phi(\cdot)$ denotes the standard normal cumulative distribution function.

The most important differences between the non-Bayesian and Bayesian approaches, other than transparency, are that: (1) the non-Bayesian approach estimates the distribution of bias for each design separately, while the Bayesian approach uses shrinkage (conditional on four covariates) to minimize the overall MSE across designs. (2) The Bayesian model has greater estimation feasibility thanks to its Markov Chain Monte Carlo feature that allows more parameters to be included in the model. For example, it can covariate-adjust the variance of bias, and not just the mean; and it can account for uncertainty in the estimates of variance components, both of which entail more sources of uncertainty in predicting the probability of future bias within the ROPE of .10 standard deviations.

## 4. RESULTS

### *4.1. Descriptive summary of the bias estimates reported in the WSCs*

Table 1 presents the distribution of estimated bias for all eight designs without modeling. At each level of analysis, both the mean and variance of estimated bias decline precipitously as design elements are added, falling to a mean of nearly zero with all three design elements and a standard deviation well under 0.10. Findings are similar for the larger sample of 50 studies that includes those where estimates of variation were not reported (results available upon request).

Table 1. Descriptive distribution of bias estimates reported in n=39 WSC studies, by observational study design

| **Pretest** | **Local** | **Rich** | **Number of observations** | **Mean** | **Standard deviation** | **Minimum** | **Maximum** |
|---|---|---|---|---|---|---|---|
| | | | *Contrast level* | | | | |
| | | | 51 | -0.48 | 0.75 | -2.65 | 0.41 |
| | | ✓ | 38 | -0.04 | 0.44 | -1.07 | 1.58 |
| | ✓ | | 38 | 0.11 | 0.24 | -0.62 | 0.42 |
| ✓ | | | 177 | -0.01 | 0.21 | -0.53 | 0.81 |
| | ✓ | ✓ | 21 | 0.12 | 0.40 | -0.35 | 1.68 |
| ✓ | | ✓ | 57 | 0.11 | 0.40 | -0.50 | 1.89 |
| ✓ | ✓ | | 39 | 0.03 | 0.16 | -0.37 | 0.43 |
| ✓ | ✓ | ✓ | 52 | 0.00 | 0.09 | -0.16 | 0.21 |

| Pretest | Local | Rich | Number of observations | Mean | Standard deviation | Minimum | Maximum |
|---|---|---|---|---|---|---|---|
| | | | *Hypothesis level* | | | | |
| | | | 17 | -0.34 | 0.55 | -2.10 | 0.19 |
| | | ✓ | 8 | 0.17 | 0.61 | -0.36 | 1.58 |
| | ✓ | | 19 | 0.06 | 0.24 | -0.62 | 0.36 |
| ✓ | | | 26 | 0.00 | 0.15 | -0.29 | 0.33 |
| | ✓ | ✓ | 7 | 0.15 | 0.42 | -0.26 | 1.05 |
| ✓ | | ✓ | 10 | 0.09 | 0.22 | -0.10 | 0.49 |
| ✓ | ✓ | | 17 | -0.01 | 0.15 | -0.29 | 0.21 |
| ✓ | ✓ | ✓ | 17 | 0.00 | 0.06 | -0.12 | 0.09 |
| | | | *Study level* | | | | |
| | | | 11 | -0.47 | 0.64 | -2.10 | 0.17 |
| | | ✓ | 5 | 0.06 | 0.24 | -0.25 | 0.31 |
| | ✓ | | 11 | 0.05 | 0.22 | -0.38 | 0.28 |
| ✓ | | | 17 | -0.02 | 0.13 | -0.22 | 0.27 |
| | ✓ | ✓ | 3 | 0.13 | 0.06 | 0.06 | 0.19 |
| ✓ | | ✓ | 8 | 0.07 | 0.22 | -0.10 | 0.43 |
| ✓ | ✓ | | 10 | 0.00 | 0.14 | -0.21 | 0.17 |
| ✓ | ✓ | ✓ | 10 | 0.01 | 0.06 | -0.08 | 0.08 |

*4.2. Meta-analysis results*

Table 2 presents the Bayesian results. As in the descriptive analysis, at all levels both mean bias and its standard deviation decrease as design elements are added, so that the best-performing design has all three design elements. Compared to the raw data, the Bayesian model generates higher estimates of the variation of bias and of all results dependent on this.

In the non-Bayesian analyses, average bias and its standard deviation also tend to decrease as design elements are added, but the standard deviations are about half of those in the Bayesian analysis and closer to the raw data but somewhat smaller. For the design with all three design elements, all standard deviations in the non-Bayesian analysis are under 0.10, and the probability of bias is 86 percent, 99 percent, and 100 percent at the contrast, hypothesis and study levels. Using the model with and without covariates makes little difference, and the t values reported here are without them.

Table 2: Meta-regression estimates from Bayesian and non-Bayesian analyses

| Pre-test | Lo-cal | Rich | Bayesian analysis | | | Non-Bayesian analysis (no covariate adjustment) | | |
|---|---|---|---|---|---|---|---|---|
| | | | Mean Bias | SD of Bias | Prob. Bias < 0.10 SDs | Mean Bias | SD of Bias | Prob. Bias < 0.10 SDs |
| *Contrast level* | | | | | | | | |
| | | | -0.10 | 0.61 | 17 | -0.47 | 0.80 | 8 |
| | | ✓ | -0.06 | 0.43 | 24 | -0.01 | 0.35 | 23 |
| | ✓ | | 0.02 | 0.35 | 28 | 0.06 | 0.22 | 34 |
| ✓ | | | -0.03 | 0.24 | 39 | -0.02 | 0.18 | 42 |
| | ✓ | ✓ | -0.02 | 0.26 | 35 | -0.06 | 0.39 | 20 |
| ✓ | | ✓ | -0.02 | 0.22 | 43 | -0.04 | 0.20 | 38 |
| ✓ | ✓ | | 0.00 | 0.18 | 49 | 0.00 | 0.11 | 65 |
| ✓ | ✓ | ✓ | -0.01 | 0.13 | 59 | 0.01 | 0.07 | 86 |
| *Hypothesis level* | | | | | | | | |
| | | | -0.10 | 0.31 | 26 | -0.47 | 0.47 | 10 |
| | | ✓ | -0.06 | 0.22 | 37 | -0.01 | 0.18 | 42 |
| | ✓ | | 0.02 | 0.22 | 38 | 0.06 | 0.15 | 46 |
| ✓ | | | -0.03 | 0.14 | 55 | -0.02 | 0.11 | 63 |
| | ✓ | ✓ | -0.02 | 0.17 | 47 | -0.06 | 0.37 | 21 |
| ✓ | | ✓ | -0.02 | 0.12 | 61 | -0.04 | 0.02 | 99 |
| ✓ | ✓ | | 0.00 | 0.11 | 64 | 0.00 | 0.05 | 96 |
| ✓ | ✓ | ✓ | -0.01 | 0.10 | 69 | 0.01 | 0.04 | 99 |
| *Study level* | | | | | | | | |
| | | | -0.10 | 0.28 | 29 | -0.47 | 0.47 | 10 |
| | | ✓ | -0.06 | 0.19 | 42 | -0.01 | 0.09 | 75 |
| | ✓ | | 0.02 | 0.19 | 43 | 0.06 | 0.15 | 46 |
| ✓ | | | -0.03 | 0.10 | 67 | -0.02 | 0.09 | 70 |
| | ✓ | ✓ | -0.02 | 0.14 | 54 | -0.06 | 0.37 | 21 |
| ✓ | | ✓ | -0.02 | 0.08 | 74 | -0.04 | 0.02 | 100 |
| ✓ | ✓ | | 0.00 | 0.08 | 77 | 0.00 | 0.05 | 96 |
| ✓ | ✓ | ✓ | -0.01 | 0.06 | 83 | 0.01 | 0.03 | 100 |

Figure 1 summarizes bias results at the preferred hypothesis-level. The pattern of relative bias shows that each design element contributes to reducing the mean and variation of bias; designs with more design elements generally perform better; and the best single design has a local comparison group, a pretest, and rich covariates. In terms of absolute bias, the mean is consistently close to zero and the standard deviation of bias is never greater than 0.10 standard deviations. However, the probability of bias being within the ROPE of 0.l0 standard deviations differs by analysis, being 95 percent in the descriptive data, 99 percent in the non-Bayesian work, but only 69 percent in the Bayesian analyses.

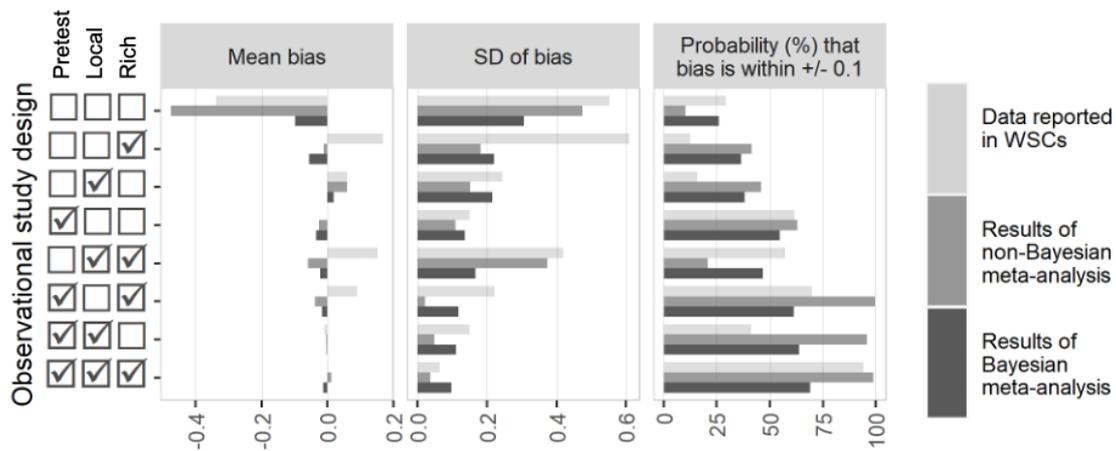

Figure 1. Risk of bias at the hypothesis-level by design. The first two panels show the estimated mean and standard deviation of bias, and the third shows the estimated probability of bias being within 0.10 standard deviations. For each observational study design, each panel shows estimates for descriptive analyses, the Bayesian meta-analysis, and the non-Bayesian meta-analysis.

## 5. DISCUSSION

This paper estimated design differences in relative and absolute bias. All analyses of relative bias agree that the superior observational study design in terms of both average bias and the lowest variation in bias is the design that combines a local comparison group, a pretest, and rich covariates. Researchers or policy-makers considering these eight options have a clear preference.

The study results for absolute bias are clearly patterned but more contingent. In the best design, all estimates of maximal and minimal bias were under .10 standard deviations at the study level (range = -.08 and .08), and the same was largely true at the hypothesis level (range = -.012 and .09). But the contrast level estimates were not so densely distributed (range = -.021 to .016). When we turn to standard deviation estimates, in the non-Bayesian model they were all under .05 standard deviations at both the study and hypothesis levels, while in the Bayesian work they were all between .05 and .10. Again, though, the covariate level was an exception. When we consider the probability of bias in the next random study being less than the ROPE, all the relevant estimates exceeded 85 percent in the non-Bayesian and in the Bayesian study-level tests. However, in the Bayesian contrast- and hypothesis-level tests the corresponding probabilities were 59 and 69 percent. Thus, while all study-level and all non-Bayesian analyses show that the best design has an acceptably high probability of tolerable future bias, there are two exceptions -- the contrast level in general and the hypothesis level in the single circumstance where the estimate of future probability of tolerable bias is 69 percent in the Bayesian results as against 99 percent in the non-Bayesian results. This pattern of absolute bias results invites the question: Are some levels and analyses of bias more appropriate than others?

      The contrast level is most prone to random error since many contrasts result from ex post facto treatment heterogeneity tests over different populations, settings, times, or ways of implementing the intervention, measuring the outcome, and analyzing the data. In high quality primary research, contrast-level tests are labelled "exploratory", and test statistics require corrections for the error rate that so many tests lead to. That is why previous meta-analyses in this series explicitly down-played contrast level results (Chaplin et al., 2018; Coopersmith et al., in press), and it is also why the range of contrasts was initially truncated in this study by excluding contrasts from appendices or that reflect minor analytic variations. Notwithstanding, policy-makers interested in the validity of unplanned contrast tests will not be heartened by the results presented here

relative to the ROPE standard. Of course, in the multiple contrast context, they cannot rely on RCTs for appropriate outcome tests either. Here, the median was 11 contrasts per study.

The study level is the most reliable of the three levels studied. But its results are potentially threatened by mis-estimated bias (meta-bias) due to averaging over heterogeneous true bias differences across all the outcomes sampled. Here. the median was 1.7 outcomes per study, and the modal dual outcomes were reading and math scores. So, with mean bias as close to zero as here, we would have to postulate that any true bias affecting reading exactly counter-balances true bias in the opposite direction for math. How plausible is this when the two scores are typically correlated about .75, and we could find no literature on design elements affecting bias in different directions over outcomes? Does an explanation based on random error around a mean of zero seem more parsimonious? In any event, policy-makers wishing to learn how biased studies with all three design elements are should be heartened by the present study-level results if they are also willing to assume that study-level variation in bias is not due to meta-bias from aggregating over true heterogeneity in the outcomes examined.

The hypothesis level corresponds most closely with normative hypothesis-formulation in basic research, and it also best balances the trade-off between precision and meta-bias. The absolute bias results at this level indicated tolerable bias over both the Bayesian and non-Bayesian analyses, but with one exception. For the estimate of future bias, the probability of bias falling within the ROPE was predicted to be 69 percent in the in the Bayesian test as opposed to 99 percent in the non-Bayesian test. Why was this?

We conducted three types of analysis -- descriptive, Bayesian and non-Bayesian. They differ in many ways, but two stand out. One is the target population for generalization. While each analysis depends on the same sample of 39 WSCs, they extrapolate differently from it. The descriptive analyses are historical and generalize only to the population of 39. The random effects model used in the non-Bayesian analyses adds uncertainty from study differences in mean impact. And the Bayesian model adds further uncertainty due to the within-study distribution of bias estimates. To which of these populations do we want to generalize? Stakeholders leery about mis-specified generalization seek to avoid or minimize extrapolation, and they may prefer to stick as close to the original data as possible. For them, the descriptive results are most important. Here, the bias maxima and minima hardly exceeded .10 standard deviations at the hypothesis- and study-levels. However, stakeholders who are willing to extrapolate but seek maximal caution about making claims from the past about the future should prefer to generalize to the most uncertain future, given that the future is never exactly like the past. They should then place most weight on the more uncertain Bayesian results. But here the within-study distribution of bias is mostly due to differences at the contrast rather than the hypothesis level, given that there are only 1.7 hypotheses per study and some studies have no hypothesis-level internal variation at all. Inevitably, the contrast level is the most uncertain and most likely to lead to higher estimates of the variation in bias. But any discussion of generalizing bias estimates is conditional on the same opportunistically chosen 39 WSCs with an opaque correspondence to any population of relevance to science and policy-making. All the interventions and outcomes deal with public policy in some way, but 60 percent of them deal with improving reading or math -- hardly reflecting a meaningful population representing public policy studies writ large. Justifying the population of available studies is a perennial meta-analytic problem, and a larger WSC sample segmented into finer-grained policy domains would better elucidate the present findings. It might also reduce the between-study differences in within-study bias variation that lowered probability estimates in the Bayesian analyses presented.

The second difference between the Bayesian and non-Bayesian analyses involves how they deal with the potential for meta-bias due to data being borrowed across observational study designs. The non-Bayesian analysis eliminates this meta-bias possibility by analyzing each design cell independently of the others. The Bayesian analysis primarily aims to increase precision, in part by borrowing data between design cells that vary in both the number and composition of studies they include -- from three to 17. The potential for meta-bias stems from such borrowing. Of course, the Bayesian model we tested does not naively shrink all design cell estimates towards a common center. Rather, it shrinks cell each towards "neighboring" cells with the same number of elements and with the greatest overlap in which design elements were included. While this shrinkage strategy is sensitive to meta-bias, but it will only eliminate it if the model is correct. Since this cannot be guaranteed, meta-bias is more plausible in the Bayesian than the non-Bayesian work.

The possibility of meta-bias from confounding design cells and the composition of studies within them led us here to conduct both Bayesian analyses that tilt the tradeoff between bias and precision in favor of precision and non-Bayesian analyses that tilt the tradeoff in the opposite direction. Future WSCs should rule out meta-bias differently, and preferably by design. The simplest way to do this is *to observe bias in all eight observational design cells within the same WSC* -- as opposed to each WSC including just one or two cells. So, after this meta-analysis was underway another mostly independent group of researchers began and completed six WSCs that included all eight designs. A meta-analysis of these six WSCs (Cook et al., 2021) showed that: (1) the relative bias findings reported here were largely replicated; (2) all indicators of study-level and non-Bayesian absolute bias showed that the design with all three design elements had bias levels within the same ROPE as here; and (3) the hypothesis-level bias results for this design were more consistent across levels and modes of analysis than here, with the probability of future bias falling within the ROPE being at least 90 percent, not just in the non-Bayesian analyses, but also in the Bayesian ones. So, a meta-analysis with less precision, but specifically designed to maximize balance and eliminate meta-bias, replicated the current findings and extended them by showing that final bias in the quasi-experiment with all three design elements was tolerable (as defined) in every test at the hypothesis and study levels.

**Acknowledgments**

Thanks are due to James Pustejovsky and Megha Joshi for very insightful analytic help with the non-Bayesian analyses.

**Funding**

This work was supported by National Science Foundation under grants 1544301 and 1760458 to Thomas D. Cook and funds from Mathematica, Inc.

**References: works cited**

**References for studies included in the descriptive analyses or the meta-analysis**

Wichman, C. J., and Ferraro, P. J. (2017), "A Cautionary Tale on Using Panel Data Estimators to Measure Program Impacts." Economics Letters, 151, 82-90.

Wilde, E. T., and Hollister, R. (2007), "How Close Is Close Enough? Evaluating Propensity Score Matching Using Data from a Class Size Reduction Experiment." *Journal of Policy Analysis and Management*, 26, 455-477.
34

# Supplemental Materials for:

# Absolute and Relative Bias in Eight Common Observational Study Designs: Evidence from a Meta-analysis

Jelena Zurovac, Thomas D. Cook, John Deke, Mariel M. Finucane, Duncan Chaplin, Jared S. Coopersmith, Michael Barna, and Lauren Vollmer Forrow

This work was supported by the National Science Foundation under grants 1544301 and 1760458 to Thomas D. Cook and funds from Mathematica, Inc.

These supplementary materials describe the methods used for calculating standardized bias estimates, defining designs with rich covariates, and estimating the standard error of bias.

## 1. *Standardized bias estimate*

Observational study bias $y$ is defined as the difference between the observational study and RCT impact estimates in standard deviation units:

$$y_{ijkl} = \frac{\hat{\theta}^{(Obs)}_{ijkl} - \hat{\theta}^{(RCT)}_{ijkl}}{S^{(RCT)}_{ijkl}} \qquad (A.1)$$

for contrast $i$, design $j$, study $k$, and outcome domain $l$. $\hat{\theta}^{(Obs)}$ and $\hat{\theta}^{(RCT)}$ are the impact estimates from the observational study and RCT respectively, and $S^{(RCT)}$ is the pooled standard deviation of the outcome for the RCT treatment and control groups, used to standardize bias.

If pooled standard deviation was not reported, we approximated it using strategies below. On occasion, WSCs reported standardized bias estimates, and we proceeded with those bias estimates.

For studies that reported treatment and RCT control group sample sizes and standard deviations, we approximated the pooled standard deviation as:

$$S^{(RCT)}_{ijkl} = \frac{\sqrt{\left(N^{(Trt)}_{ijkl} - 1\right)\left(S^{(Trt)}_{ijkl}\right)^2 + \left(N^{(Contr)}_{ijkl} - 1\right)\left(S^{(Contr)}_{ijkl}\right)^2}}{\sqrt{N^{(Trt)}_{ijkl} + N^{(Contr)}_{ijkl} - 2}} \qquad (A.2)$$

where $N^{(Trt)}$ and $N^{(Contr)}$ are the treatment and RCT control group sample sizes, respectively, $S^{(Trt)}$ is the standard deviation of the outcome for the treatment group, and $S^{(Contr)}$ is the standard deviation of the outcome for the RCT control group.

For the remaining studies we used one of the following strategies to approximate the standard deviation of the outcome. Depending on which information was reported, we used the first available strategy strategies, in the order below.

- For studies that reported the standard deviation of the treatment group, we approximated the pooled standard deviation with the standard deviation of the treatment group, $S^{(Trt)}$.
- For studies that reported standard errors of the RCT impact estimate ($s^{(RCT)}$), we approximated the pooled standard deviation as:

$$S_{ijkl}^{(RCT)} = \frac{s_{ijkl}^{(RCT)}}{\sqrt{\frac{1}{N_{ijkl}^{(Trt)}} + \frac{1}{N_{ijkl}^{(Contr)}}}} \quad (A.3)$$

- For studies that reported a F-statistic for the RCT impact ($F^{(RCT)}$), we approximated the pooled standard deviation as:

$$S_{ijkl}^{(RCT)} = \frac{\sqrt{F_{ijkl}^{(RCT)}}}{\sqrt{\frac{1}{N_{ijkl}^{(Trt)}} + \frac{1}{N_{ijkl}^{(Contr)}}}} \quad (A.4)$$

- For studies that reported a *p*-value from the t-test on the RCT impact, we approximated the pooled standard deviation as:

$$S_{ijkl}^{(RCT)} = \frac{t_{ijkl}^{(RCT)}}{\sqrt{\frac{1}{N_{ijkl}^{(Trt)}} + \frac{1}{N_{ijkl}^{(Contr)}}}} \quad (A.5$$

where $t^{(RCT)}$ is the t-statistic derived from the reported *p*-value with $N^{(Trt)} + N^{(Contr)} - 2$ degrees of freedom.

## *2. Definition of rich covariates*

The main text of the article describes how we defined the use of a pretest and a local comparison group. This section provides additional detail about designs with rich covariates.

We considered covariates to be rich if a study matched on or regression-adjusted for variables that covered at least four unique conceptual domains. We predefined the domains at the onset of the study. In creating these domains, our main goal was to differentiate between studies that match on or control for many versus few types of attributes of study participants.

- Basic demographics included age, gender, and race and ethnicity

- Education characteristics included education history of participants or family members, for example, children's test scores, own achievement of a degree, and mother's years of education

- Other socioeconomic characteristics included earnings, free lunch status, housing, employment, family size, and marital or cohabitation status

- Health status included self-reported health status, severity of illness, diagnoses and disability status

- Health utilization included utilization of health care services such as use of the emergency room
- Health expenditures included expenditures for health care services
- Participation in other programs or initiatives that might influence selection into treatment or outcomes included variables such as participation in Supplemental Nutrition Assistance Program (SNAP) for job training interventions.
- Geographic characteristics other than those covered by other domains included population density, whether units are in rural or urban areas, and resources per capita (such as number of hospitals per capita).
- Behavioral and emotional characteristics of participants included information such as motivation
- Other domain included characteristics that did not belong in other categories, for example, whether school is a charter school, availability of resources (such as running water), and crime statistics

## *3. Estimating the standard error of bias*

The main contribution of this paper is that we estimate the variation in bias from observational studies. The key challenge in doing so is distinguishing between variability in bias estimates due to noise (that is, random estimation error) from variation in signal (that is, variation in observational study bias). We estimated observational study bias, signal, by accounting for the variation due to sampling error, noise, using standard errors of the bias estimates. Because not all studies reported these standard errors and we needed a consistent measure across studies and estimates of bias, we estimated the standard error of bias estimates by taking into account that RCT and observational study share a treatment group. We use the inverse of the estimated standard error of the bias as a weight in our analyses.

Equation (A.1) from the previous section, for calculating bias estimates, implies that:

$$\left(s_{ijkl}^{(y)}\right)^2 = \left(s_{ijkl}^{(Obs)}\right)^2 + \left(s_{ijkl}^{(RCT)}\right)^2 - 2c\left(\hat{\theta}_{ijkl}^{(Obs)}, \hat{\theta}_{ijkl}^{(RCT)}\right) \tag{A.6}$$

where for contrast $i$, design $j$, study $k$, and outcome domain $l$:

- $c\left(\hat{\theta}_{ijkl}^{(Obs)}, \hat{\theta}_{ijkl}^{(RCT)}\right)$ is the covariance between observational and RCT effect estimates due to sampling, and
- $\left(s_{ijkl}^{(y)}\right)^2, \left(s_{ijkl}^{(Obs)}\right)^2,$ and $\left(s_{ijkl}^{(RCT)}\right)^2$ are the squared standard errors for the bias estimate, the observational study, and RCT estimated effects, respectively.

Our estimate of $\left(s_{ijkl}^{(y)}\right)^2$ accounts for the covariance term in equation (A.6) using the sample sizes of the treatment, control, and comparison groups and is presented in equation (A.13) below. Our approximation can be justified as follows. Suppose that the RCT and observational studies were both based on simple comparisons of outcomes that had standard deviations of one in the treatment, control, and comparison groups and there was no clustering or use of matching or control groups. In this case the variances of the means of the treatment, control,

(A.8

(A.9

and comparison groups would be $1/N_{ijkl}^{(Trt)}$, $1/N_{ijkl}^{Contr(Contr)}$, and $1/N_{ijkl}^{(Comp)}$, where $N_{ijkl}^{(Trt)}$ is the treatment group sample size, $N_{ijkl}^{(Contr)}$ is the control group sample size, and $N_{ijkl}^{(Comp)}$ is the comparison group sample size. In this case:

$$\left(s_{ijkl}^{(RCT)}\right)^2 = \frac{1}{N_{ijkl}^{(Trt)}} + \frac{1}{N_{ijkl}^{(Contr)}}$$

$$\left(s_{ijkl}^{(Obs)}\right)^2 = \frac{1}{N_{ijkl}^{(Trt)}} + \frac{1}{N_{ijkl}^{(Comp)}}$$

and

$$c(Obs_{ijkl}, RCT_{ijkl}) = \frac{1}{N_{ijkl}^{(Trt)}} \tag{A.10}$$

This implies that:

$$\left(s_{ijkl}^{(y)}\right)^2 = \left(s_{ijkl}^{(Obs)}\right)^2 + \left(s_{ijkl}^{(RCT)}\right)^2 - 2c(Obs_{ijkl}, RCT_{ijkl})$$

$$= \left(\frac{1}{N_{ijkl}^{(Trt)}} + \frac{1}{N_{ijkl}^{(Contr)}} + \frac{1}{N_{ijkl}^{(Trt)}} + \frac{1}{N_{ijkl}^{(Comp)}} - 2\frac{1}{N_{ijkl}^{(Trt)}}\right)$$

$$= \frac{1}{N_{ijkl}^{(Contr)}} + \frac{1}{N_{ijkl}^{(Comp)}} \tag{A.11}$$

$$\left(s_{ijkl}^{(Obs)}\right)^2 + \left(s_{ijkl}^{(RCT)}\right)^2 = \left(\frac{1}{N_{ijkl}^{(Contr)}} + \frac{1}{N_{ijkl}^{(Comp)}} + \frac{2}{N_{ijkl}^{(Trt)}}\right) \tag{A.12}$$

and

$$\left(s_{ijkl}^{(y)}\right)^2 = \left(\left(s_{ijkl}^{(Obs)}\right)^2 + \left(s_{ijkl}^{(RCT)}\right)^2\right) \left(\frac{\frac{1}{N_{ijkl}^{(Contr)}} + \frac{1}{N_{ijkl}^{(Comp)}}}{\frac{1}{N_{ijkl}^{(Contr)}} + \frac{1}{N_{ijkl}^{(Comp)}} + 2\frac{1}{N_{ijkl}^{(Trt)}}}\right) \tag{A.13}$$

Thus, the sample sizes in equation (A.13) adjust $\left(s_{ijkl}^{(Obs)}\right)^2 + \left(s_{ijkl}^{(RCT)}\right)^2$ down by the correct amount to approximate $\left(s_{ijkl}^{(y)}\right)^2$ given the assumptions listed above. Since those assumptions do not hold for most bias estimates, this formula is an approximation.